\documentstyle[12pt]{article}
\begin{document}
\newcommand{\beq}{\begin{equation}}
\newcommand{\eeq}{\end{equation}}
\newcommand{\KS}{\int D\xi^\perp}
\newcommand{\KSa}{\int D\xi^{a\perp}}
\newcommand{\KSm}{\xi_\mu^\perp}
\newcommand{\KSn}{\xi_\nu^\perp}
\newcommand{\OSn}{\omega_\nu^\perp}
\newcommand{\OSm}{\omega_\mu^\perp}
\newcommand{\mn}{\mu\nu}
\newcommand{\Nn}{\nabla_\nu}
\newcommand{\Nm}{\nabla_\mu}
\newcommand{\Nr}{\nabla_\rho}
\newcommand{\Nk}{\nabla^2}
\newcommand{\Pm}{\perp\mu}
\newcommand{\RA}{\rightarrow}
\newcommand{\nonumsection}[1] {\vspace{12pt}\noindent{\tenbf #1}
        \par\vspace{5pt}}
\newcommand{\bibit}{\nineit}
\newcommand{\bibbf}{\ninebf}
\renewenvironment{thebibliography}[1]
\
                               {\ninerm
         \baselineskip=12pt
\

         \begin{list}{\arabic{enumi}.}
        {\usecounter{enumi}\setlength{\parsep}{0pt}
         \setlength{\leftmargin 17pt}{\rightmargin 0pt}

         \setlength{\itemsep}{0pt} \settowidth

\

        {\labelwidth}{#1.}\sloppy}}{\end{list}}

\newcommand{\RR}{\mu\nu\rho\sigma}
\thispagestyle{empty}
\baselineskip=0.6cm
\begin{flushright}{TIT/HEP-257}\\
{May 1994}\end{flushright}
\begin{center}
\vspace{15mm}

{\large\bf The  finite vacuum energy for spinor, scalar}\\
{\large\bf and vector fields}\\

\bigskip
\vspace{0.3in}

{N.~N.~Shtykov}\\
\medskip{\it Department of Physics , Tokyo Institute of
Technology},\\
{\it Oh-Okayama, Tokyo, 152, Japan}$^{\dagger}$\\

\end{center}
\vspace{15mm}

\centerline{\bf Abstract}
\begin{quotation}
We compute the one-loop potential (the Casimir energy) for
scalar, spinor and vectors fields on the spaces $\,R^{m+1}\,
\times\,Y$ with $\,Y=\,S^N\,,CP^2$. As a physical model we
consider spinor electrodynamics on four-dimensional product
manifolds. We examine the cancelation of a divergent part of
the Casimir energy on even-dimensional spaces by means of
including the parameter $\,M$ in original action. For some
models we compare our results with those found in the literature.
\end{quotation}

\vfill
\noindent
$^{\dagger}$ On leave from: {\em Irkutsk university, Russia};\\
Electronic mail: shtykov@phys.titech.ac.jp

\newpage
\setcounter{page}{2}
{\bf 1.Introduction}\\

The study of vacuum fluctuations in quantum field theory has been a
subject of extensive research$^1$. The Casimir energy may be
defined as the energy due to the distortion of the vacuum. For
systems with a finite number of degrees of freedom the vacuum energy
is finite and measurable. However, for many models in quantum
field theory, the vacuum energy  has a divergent value and some
regularization
procedure should be used to remove  this divergence. A
meaningful definition of the physical vacuum energy must take
into account the fact that quantum fields always exist in interaction
with various external fields. In a typical situation, a field is
given in the presence of macroscopic objects or it is confined
to a finite cavity. The field then has to satisfy certain
boundary conditions. In such cases one can treat the vacuum
energy as a function of a suitable set of parameters $\,{\gamma\,
}$ which can characterize the given geometrical configuration.
Hence , we have to calculate the physical vacuum energy
( Casimir energy ) of a quantized field with respect to its
interaction with the external configuration. It can be defined as
the difference between the energy of the distored vacuum configuration
and that of the free vacuum configuration .
 The original effect considered by Casimir$^2$ was the
attraction of two uncharged parallel conducting plates at zero
temperature.
 Of particular physical interest is the vacuum energy of the
electron-positron field in strong electromagnetic fields which
induces an electron-positron pair to be produced $^3$. The presence of
boundary effects has been central to investigate the vacuum
energy in bag models of QCD$^4$. The Casimir effect is an essential
ingredient in multidimensinal theories of unified interactions.
For example, it has been shown that the Casimir energy in
a five-dimensinal Kaluza-Klein model with one compact dimension
tends to shrink the size of the compact dimension$^5$.

In order to
have a well-defined notion of vacuum energy it is convenient to
use an ultrastatic space-time $\ R^1\ \times \ M^d \;$ where  $
M^d \;$ is some Riemann manifold. In a number of works
the one-loop effective potential ( directly related with the
Casimir energy) was calculated for scalar and
spinor fields with a background geometry of $\,R^4\ \times\  S^N\,$.
The
effective potential has the structure $\,C_N\,\rho^{-4}\,$ where
$\,C_N\,$ is a numerical coefficient, $\,\rho\,$ is a radius of
$\,S^N\,$ . Results obtained for odd $N\,$ are finite for both
scalars and spinors$^6$. In the case of an even $N\,$ it has been
found$^7$ that for scalars and spinors the coefficient
$\,C_N\,$
has a divergent part, which is expressed in terms of geometrical
invariants of $\,S^N\,$. It effectively describes high-energy
states of the theory. A finite part of $\,V^{(1)}\,$ is connected
with both high and low-energy behaviours of the theory and
depends on topological properties of the manifold.
Consider the Casimir energy ($\,h\,,\,c\,=1\,$)
\beq
E_c\ = \frac{1}{2}\,\sum \omega_n.
\eeq
Here $\omega^{2}_n\,$ are eigenvalues of the Hamiltonian of some vacuum
configuration. We can regularize the divergent sum (1) by
defining
\beq
E^{(reg)}_c = \frac{1}{2}\,\mu^{\delta}\,(\omega^{2}_n\,)^{ \frac{1}{2}-
\delta}\, = \frac{1}{2}\,\mu^{\delta}\,\zeta_n\,(-  \frac{1}{2} + \delta
).
\eeq
A scaling parameter $\,\mu\,$ is introduced in order to keep
 the $\,\zeta\,$- function dimensionless for all $\delta\,$. In
many interesting models $\,E^{(reg)}_c\,$ can have a divergent
part, associatied with a pole of $\,\zeta_n\,(- \frac{1}{2} +
\delta)\,$. In these cases, we should make the renormalization
$\,E^{(reg)}_c\,$ to remove the divergent term in (2) . However,
in a wide class of models the calculation of quantum
effects has been limited by one-loop approximation. In such
models, the way in which divergences are removed is not unique.
If we have no strong physical restrictions an ambiguity is
left. The renormalization scheme used in Ref. 8
adopts the minimal substraction scheme which is equivalent to
simply removing the pole
\beq
E^{(ren)}_c = \frac{1}{2}\,\mu\,\lim_{\delta\,{\RA}
0}(\,\zeta_n\,(-  \frac{1}{2} + \delta) + \zeta_n\,(- \frac{1}{2} - \delta))
.\eeq

In this article we calculate the one-loop effective potential
for spinor, scalar and vector fields on $\,R^m\;\times\;S^N\, ,
R^m\;\times\; CP^2\,$ where $\,R^m\,$ is  the $\,m$-dimensional Euclidean
space,
$\,S^N\,$ is the N-dimensional sphere, $\,CP^2\,$ is the projective space.
We discuss another approach to remove the
divergent part of the vacuum energy in these multidimensional
models. For the original action we include the parameter $\,M\,$,
which may be
treated as an  effective value appearing after the reduction of
some of the more
generalized theory. The parameter $\,M\,$ should be interpreted as a
scale, which includes the effects associated with boundaries,
curvature and masses.
It will be shown that in a number of models under consideration
here a
divergence in the vacuum energy can be removed by a suitable
choice of $\,M\,$. In this case there is no  ambiguity in the
calculation of the finite part of the energy. \\

{\bf 2. Analytic regularization of the one-loop potential} \\

It is easy to observe that the expression
$\;\sum_{(k)}\lambda_k\;$
connected with the calculation of $\;det\,\Delta\;$ where
$\;\lambda_k\;$ are (discrete or continuous) eigenvalues of
operator $\,\Delta\;$, is ill-defined and some kind of
regularization should be used. We shall consider models on the
spaces $\,R^{m+1}\ \times\  Y\,$ where $\,Y\,$ is  compact with a
discrete spectrum $\;\lambda_n\;$ of a second-order elliptic
operator $\,\tilde\Delta\;$. To give a finite definition for
$\;det\,\Delta\;$ we will use  $\,\zeta\,$- regularization.
Here, the regularized one-loop effective potential related with
$\,\tilde\Delta\;$ can be written as $^6$
 $$
V^{(1)} = \lim_{s\RA 1}\;\frac{\mu^{2\epsilon}}{2 (2\pi^m)}\,\int
d^mk\,\sum_{n}d_{n}{\lambda_{n,k}}^{s/2}
$$ where $\,\epsilon\,=\,(1-s)/2\,$.
Taking into account that $\;\lambda_{n,k} = \lambda_n + k^2\;$
and using the formula
$$\int d^mk\,(k^2 + y)^{s/2} =
\frac{\pi^{m/2}\Gamma{((-s-m)/2)}}{\Gamma{(-s/2)}}\,y^{(m+s)/2}$$
we get
\beq
 V^{(1)} = \lim_{s\RA 1}\;\frac{\Gamma{((-s-m)/2)}\,\mu^{2\epsilon}}{2
(4\pi)^{m/2}\,\Gamma(-s/2)}\,\sum_{n}d_{n}{\lambda_{n}}^{(m+s)/2}.
\eeq
The potential (4) just gives the Casimir energy density
$$ E^{reg}_c\,= V^{(1)}\Omega^{-1}(R^m) $$
where $\,\Omega(R^m)\,$ is the volume of $\,R^m$.
The eigenvalues $\;\lambda_n\;$ and their multiplicities
$\;d_{n}\;$ can be obtained by the harmonic expansion method$^9$ on
an arbitrary homogeneous space $\;Y = G/H\;$ . We should
define a set of irreducible representations (irreps) of $\;G\;$
giving the needed irrep of $\;H\;$ under reduction
$\;G\downarrow H\;$ . As an explicit example in Sect. (6) we
compute $\;\lambda_n\;$ and $\;d_{n}\;$ for scalars and vectors
on $\;CP^2\;$ . For a second-order operator $\,\tilde\Delta\;$ we
have
\beq
\lambda_n = a_2\,n^2 + a_1\,n + a_0,\ \ \ \ \ \ \ d_{n} =
b_r\,n^r + ... + b_1\,n + b_0
\eeq
In fact, from (4), (5) we see that the main quantity we need to
compute is
\beq
\lim_{\epsilon\RA 0}\;\mu^{2\epsilon}\,\zeta(n + 2\epsilon, p)\;\Gamma{(l
+\epsilon)}\,\Gamma^{-1}(\epsilon -1/2)
\eeq
for integer $\;n , l\;$ . The function $\;\Gamma{(l
+\epsilon)}\;$ has simple poles at points $\;l = -k\ \ (k =
0,1,2...)\;$ with the expansion$^{10}$
\beq
\Gamma{(-k +\epsilon)}\;= \frac{(-1)^k}{k!}\,(\frac{1}{\epsilon}
+ \psi(k +1) + O(\epsilon)).
\eeq
The function $\,\zeta(n + 2\epsilon, p)\;$ has a simple pole at
$\,n=1\,$
\beq
\zeta(n + 2\epsilon, p) = \frac{1}{2\,\epsilon} - \psi(p) +
O(\epsilon).
\eeq
The most general form of (6) with (7),(8) at $\;\epsilon\RA 0\,$
is
\beq
A_{n,l}\,\frac{1}{\epsilon^2} + B_{n,l}(p)\,\frac{1}{\epsilon} +
F_{n,l}(p) + O(\epsilon).
\eeq
The nonzero coefficients $\,A_{1,-k} = (-1)^{k+1}(4\,\pi^{1/2}k!)^{-1}\,$ do
no\
t
appear in our models.  The coefficients
\beq
B_{1,k}(p)\;=\,- \frac{1}{4\pi^{1/2}}\,\Gamma(k)\ \ \ \ B_{n,-k}(p)\;=
 \frac{(-1)^{k+1}}{2\pi^{1/2}\,k!}\,\zeta(n, p)
\eeq
reflect the local properties of manifolds and are constructed
from the curvature tensor and its contractions. The objects
$$
F_{1,l}(p)\;=\,\frac{\Gamma(l)}{4\,\pi^{1/2}}\,(2\psi(p)\, -
\,\psi(l)\,+\,\psi(- \frac{1}{2})\,-2\log\mu)$$
\beq F_{n,-k}(p)\;=\frac{(-1)^{k+1}}{2\pi^{1/2}\,k!}\,( \frac
{d\,\zeta(n + 2\epsilon,
p)}{d\,\epsilon} + (\,\psi(k +1)\,-\,\psi(- \frac{1}{2})\,+2\log\mu)\zeta(n,
p)\
)
\eeq
 contain topological and geometrical information
about the spaces $\,R^{m+1}\,\times\, Y\,$ . It should be noted
that analytic and numerical values of $\,B_{n,l}(p)\;$ have been
obtained in many models from asymptotic expansion of the heat
kernal$^{7,11}$.

It may be shown from general constructions$^6$ that
$\,B_{n,l}(p)= 0\;$  for odd-dimensional manifolds without
boundaries. Numerical calculations confirm this result$^{6,7,12}$. In
these cases the finite vacuum energy is  defined by
$\,F_{n,l}(p)\;$. For even-dimensional manifolds due to the
presence of the divergence there is an ambiguity in the
numerical value of the finite part of $\;V^{(1)}\;$ . This
ambiguity can be removed if there are real solutions $M_i\,$ for
the equation
\beq
S(M)\,=\sum_{k=0}^L\;c_k\,M^k\;=\;0,
\eeq
where $\,S(M)\,$ is the divergent part of the one-loop potential
with the eigenvalues $\lambda_n + M\,$. Explicit examples will be considered in
Sect. 2-7. \\

{\bf 3. The one-loop potential for spinors on $\,R^{m+1}\ \times\
S^N\,$} \\

The eigenvalues and degeneracies of the squared Dirac operator on
the $\,N\,$-sphere with the radius $\,\rho\,$ are given by$^{6}$
\beq
\lambda_n \;=\, (n +\frac{ N}{2})^2\rho^{-2},\ \ \ \ \ \ \ d_{n}\;=
\,\frac{2^{[N/2]+1}\,\Gamma(n + N)}{\Gamma(N)\,n!}.
\eeq
The corresponding one-loop spinor potential on $\,R^{m+1}\,\times\;Y\;$ can
be written as
\beq
V^{(1)}_{sp}\; = \lim_{s\RA
%% FOLLOWING LINE CANNOT BE BROKEN BEFORE 80 CHAR
1}\;-\frac{2^{[(m+s)/2]}\;\Gamma{((-s-m)/2)}\,\mu^{2\epsilon}}{2(4\pi)^{m/2}\,\\
Gamma(-s/2)}\,\sum_{n}d_{n}
{\lambda_{n}}^{(m+s)/2}.
\eeq
Inserting (13) into (14) and using (9)-(11) we compute $\;V^{(1)}_{sp}(m,N)\;$
for different $\;m\,,N\;$.

For $\;m = 0\;$ all values of $\;V^{(1)}_{sp}(m,N)\;$ are finite.
$$ V^{(1)}_{sp}(0,1)\; =\; -\,\zeta(-1, \frac{1}{2}) =
-0.0417, $$
 $$ V^{(1)}_{sp}(0,2)\; =\; -2\,\zeta(-2) = 0 ,$$
 $$ V^{(1)}_{sp}(0,3)\; =\; -\,\zeta(-3,\frac{3}{2}) +
\frac{1}{4}\zeta(-1,\frac{3}{2}) = 0.0177, $$
 \beq V^{(1)}_{sp}(0,4)\; =\; -\frac{2}{3}\,(\zeta(-4) -
\zeta(-2)) = 0.
\eeq
Here $\;\zeta(v, p)\ \ \ (\zeta(v, 1)\equiv\,\zeta(v))\,$ is the
generalized zeta function and throughout the paper we set $\,\rho\,=1\,$ for
si\
mplicity.

For $\;m = 1\;$ we have the finite expressions
\beq V^{(1)}_{sp}(1,1)\; =\;
-\frac{3\,\zeta(3)}{16\,\pi^3} = -0.0073,\ \ \ \ \ \ \
  V^{(1)}_{sp}(1,3)\; =
\frac{3\,\zeta(3)}{64\,\pi^3} + \frac{45\,\zeta(5)}{64\,\pi^5} = 0.0042 \\
\eeq
We find divergencies in $\;V^{(1)}_{sp}\;$ for even $\,N\,$.
$$ V^{(1)}_{sp}(1,2)\; =\;-0.0027\frac{1}{\bar{\epsilon}} -
0.00445, $$
 \beq  V^{(1)}_{sp}(1,4)\; =\;0.0013\frac{1}{\bar
 {\epsilon}}\,+0.0018.
 \eeq
where $\,\bar{\epsilon}\equiv\,\epsilon
(1+\epsilon\,\log\mu)^{-1}$.
To cancel these divergencies we take $\;\lambda_{n}\,+ M_i\,$ instead
of $\;\lambda_{n}\,$ in (13) where $\,M_i\,$ are the solutions of (12).
To evaluate (14) we make the following expansion
\beq
\sum_{n=0}^\infty (n+q)^l((n+q)^2+M)^{\frac{m+s}{2}}\,=\,\sum_
{ k=0}^\infty\;\frac{\Gamma(k-(m+s)/2)}{\Gamma(-(m+s)/2)\Gamma(-s/2)}\,
\zeta(2k-l-m-s,\,q)\,M^k.
\eeq
Then, for $\;m = 1\;, N = 2\;$ we get
$$ \frac{1}{4}\,M^2 - \zeta(-1)\,M - \zeta(-3) = 0\ \RA $$
\beq
M_1 = \;0.0805,\ \ \ \ \ \ \ \ \ \ M_2\, = \;-0.4139
\eeq
For $\;m = 1\;, N = 4\;$
$$ \frac{1}{12}\,M^3\, +\frac{1}{4}\, M^2 + (\zeta(-3) - \zeta(-1))\,M +
\zeta(-5) - \zeta(-3) = 0 \ \RA $$
\beq
M_1\; =\,0.1038,\ \ \ \ \ \ \ \ \ \ M_2\; = \,-0.5588,\ \ \ \ \ \ \
\ \ \ M_3\; = \,-2.5450.
\eeq
Inserting $\;M_i\,$ from (19),(20) into (14) and using
(9)-(11),(18) we obtain
the finite values of $\;V^{(1)}_{sp}\;$
$$  V^{(1)}_{sp\ i}(1,2)\; = \frac{1}{\pi}\,( -0.0143\, +
0.28274\,M_i\, + 0.1443\,M^2_i\,+
 \sum^{\infty}_{k= 3}\;\frac{\zeta(2k-3)\,(-M_i)^k}{k(k -1)})\ :$$
\beq
V^{(1)}_{sp\;1}\; =\;0.0030,\ \ \ \ \ \ \ \ \ \ V^{(1)}_{sp\;2}\;
=\;-0.0283\,;
\eeq
$$ V^{(1)}_{sp\ i}(1,4) = \,\frac{1}{3\pi}\,(0.0171\,
-0.2886\,M_i -0.1860\,M^2_i +\, 0.0689\,M^3_i + $$
 $$
 \sum^{\infty}_{k= 4}\;\frac{(\zeta(2k-5)\, -\,
\zeta(2k-3))\,(-M_i)^k}{k(k -1)})\ :$$
\beq
V^{(1)}_{sp\;1}\; =\;-0.0016,\ \ \ \ \ \ \ \ \ \
 V^{(1)}_{sp\;2}\; =\; 0.0116,\ \ \ \ \ \ \ \ \ \
V^{(1)}_{sp\;3}\; =\; -0.0799.
\eeq
For $\;m = 2,\,3\;$ we have
$$ V^{(1)}_{sp}(2,1) =\,\frac{\zeta(-3,
\frac{1}{2})}{(3\,\pi)} = \,-7.74\  10^{-4 },\ \ \ \ \ \ \
V^{(1)}_{sp}(2,2) =\,0, $$
 $$  V^{(1)}_{sp}(2,3) =\,\frac{\zeta(-5,
3/2)\, - \frac{1}{4}\, \zeta(-3, \frac{3}{2})}{3\pi} = \,6.01\
10^{-4},\ \ \ \ \ \ \ V^{(1)}_{sp}(2,4) =\,0,  $$
$$  V^{(1)}_{sp}(3,1) = -
\frac{45\,\zeta(5)}{4\,(2\pi)^6} =\,-1.90\ 10^{-4},\ \ \ \ \ \ \
 V^{(1)}_{sp}(3,2) = -1.01\
10^{-4}\,\frac{1}{\bar{\epsilon}}\,-\,4.73 \ 10^{-4},
 $$
$$ V^{(1)}_{sp}(3,3)\;
=\;\frac{2835\,\zeta(7)}{8\,(2\pi)^8} \;+\,
\frac{45\,\zeta(5)}{16\,(2\pi)^6}\;=\,1.95\ 10^{-4},
$$
\beq V^{(1)}_{sp}(3,4)\; =\;6.87\
 10^{-5}\,\frac{1}{\bar{\epsilon}}\,+ \,5.66\ 10^{-5}.
\eeq
To obtain the finite $\;V^{(1)}_{sp}\;$ for $\,N = 2 \;,4\,$ we
should take the  values of $\,M_i\;$ removing the divergencies
in (23)
$$ N = 2,\ \ \ \ -\frac{1}{12}M^3\, - \frac{1}{24}M^2
 +\,\zeta(-3)M +\,\frac{1}{2}\zeta(-5) =\,0, \,\RA\
 M = \,-0.69365\,; $$
 $$ N = 4, \ \ \ \ \frac{1}{48}M^4 +\,\frac{1}{12}M^3
 +\,\frac{11}{240}\,M^2 +
  (\zeta(-5) -\,\zeta(-3))\,M\;+$$
 \beq \frac{1}{2}\,\zeta(-7) - \,
 \zeta(-5) =\,0,\, \RA\
 M_1\;=\, -3.26525,\ \ \ \ M_2\;=\, -0.9952.
\eeq
Inserting (24) into (14) and making a straightforward calculation
we get
\beq
V^{(1)}_{sp}(3,2)\; =\;-0.017, \ \ \ \ \ \ \ \ \
V^{(1)}_{sp\;1}(3,4)\; =\;-0.0161,\ \ \ \ \ \
\ \ \ V^{(1)}_{sp\;2}(3,4)\; =\;0.0015.
\eeq

It should be noted that the results (23) coincide with those
obtained in Ref.7 for $\,R^4\,\times\,S^N$ except for the finite parts
of $\;V^{(1)}_{sp}\;$ when
$\,N = 2 \;,4\,$. We shall shortly discuss this difference  in Sect.8.
\\

{\bf 4. The one-loop potential for scalars on $\,R^{m+1}\
\times\
S^N\,$} \\

The one-loop scalar potential on $\,R^{m+1}\,\times\;Y\;$ has
 the same form as (4). The eigenvalues and degeneracies of the
Laplace operator on the $\,N\,$-sphere are given by$^6$
\beq
\lambda_n \;=\, n(n\,+\,N\,-\,1),\ \ \ \ \ \ \ d_{n}\;=\,\frac
{ (2n \,+\,N\, - 1)\,\Gamma(n\, +\,N\, -1)}{\Gamma(N)\,n!}
\eeq
Substituting (26) in (4) and taking into account (9)-(11),(18) we find the
following numerical values of $\;V^{(1)}\;$ .
The divergent parts of $\;V^{(1)}\;$ are
$$  V^{(1)}_{div}(0,3)
= -0.03125 \frac{1}{\bar{\epsilon}},\ \ \ \ \ \ \
 V^{(1)}_{div}(1,2) = -0.00265 \frac{1}{\bar{\epsilon}}, \ \ \ \ \ \ \
\ V^{(1)}_{div} = -0.0078 \frac{1}{\bar{\epsilon}} $$
 \beq
  V^{(1)}_{div}(2,3) = -8.29\
10^{-4}\frac{1}{\bar{\epsilon}}, \ \ \ \ \
 V^{(1)}_{div}(3,2) = -4.02 \ 10^{-5}\frac{1}{\bar{\epsilon}}, \ \ \ \ \ \ \ \
 V^{(1)}_{div}(3,4) = -2.50\ 10^{-4}\frac{1}{\bar{\epsilon}}
\eeq
The values of the finite parts of $\;V^{(1)}\;$ are
$$ m\ \ \ \ \ \ \ \ \ N\,=\,1\ \ \ \ \ \ \ \ \ \ \ \ N\,=2 \ \ \ \ \ \ \ \ \
\ \ \ \ \ \ \ \ N\,=\,3\ \ \ \ \ \ \ \ \ \ \ \ \ \ \ N = 4$$
$$ 0\ \ \ \ \ \ \ \ \ -0.0833\ \ \ \ \ \ \ \ \ -0.1325\ \ \ \ \ \ \
\ \ \ \ \ \ -0.2057\ \ \ \ \ \ \ \ \ -0.2159 $$
 $$ 1\ \ \ \ \ \ \ \ \ -0.0048\ \ \ \ \ \ \ \ \ -0.0117\ \ \ \ \ \ \
\ \ \ \ \ \ -0.0115\ \ \ \ \ \ \ \ \ -0.0218 $$
$$ 2\ \ \ \ \ \ \ \ -4.42\ 10^{-4}\ \ \ \ \ \ \ -4.64\ 10^{-4}\ \
\ \ \ \ \ \ -0.0013\ \ \ \ \ \ \ -6.48\ 10^{-4}$$
\beq 3\ \ \ \ \ \ \ \ -5.06\ 10^{-5}\ \ \ \ \ \ \ -9.10\ 10^{-5}\ \
\ \ \ \ \ \ 7.58 \ 10^{-5} \ \ \ \ \ \ \ -2.48\ 10^{-5}
\eeq
We look for real solutions of the
 equations (12) to remove divergencies  (26) and to compute the
finite one-loop potential . We have for $\;m = 0\;, N = 3\;$
\beq
M\;=\;1,\ \ \ \ \ \ \ \ \ \ V^{(1)}_{fin}(0,3)\;=\;0.0042.
\eeq
For $\;m = 1\;, N = 4\;$
 $$ \frac{1}{12}\,(\frac{9}{4} - M)^3 -
\frac{1}{16}\,(\frac{9}{4} - M)^2 + $$
 $$ (\zeta(-3, \frac{3}{2}) -\,\frac{1}{4}\,\zeta(-1,
\frac{3}{2}))\,(\frac{9}{4} -\,M\,)\,+ \frac{1}{4}\,\zeta(-3,
\frac{3}{2}) -\,\zeta(-5, \frac{3}{2})\,= 0\ \RA  $$
\beq
M\;=\;1.2272,\ \ \ \ \ \ \ \ \ \ V^{(1)}_{fin}(1,4)\;=\;0.0129.
\eeq
For $\;m = 2\;, N = 3\;$
\beq
M\;=\;1,\ \ \ \ \ \ \ \ \ \ V^{(1)}_{fin}(2,3)\;=\;1.05\ 10^{-4}.
\eeq
For $\;m = 3\;, N = 2\;$
$$ \frac{1}{6}\,(\frac{1}{4} - M)^3\,+\,\zeta(-1,
\frac{1}{2})\,(\frac{1}{4} - M)^2\,- 2\zeta(-3,
\frac{1}{2})\,(\frac{1}{4} - M)\,+\,\zeta(-5,
\frac{1}{2})\;=\,0\ \RA $$
\beq
M\;=\;0.5077,\ \ \ \ \ \ \ \ \ \ V^{(1)}_{fin}(3,2)\;=\;9.68\ 10^{-6}.
\eeq
For $\;m = 3\;, N = 4\;$
$$ \frac{1}{48}\,(\frac{9}{4} - M
)^4\,-\,\frac{1}{48}\,(\frac{9}{4} - M )^3\,
+\,\frac{1}{2}\,(\zeta(-3, \frac{3}{2})
-\,\frac{1}{4}\,\zeta(-1,
\frac{3}{2}))\,(\frac{9}{4} -\,M\,)^2\; + $$
$$ (\frac{1}{4}\,\zeta(-3, \frac{3}{2}) -\,\zeta(-5,
\frac{3}{2}))(\frac{9}{4} -\,M\,)\, -\frac{1}{8}\,\zeta(-5,
\frac{3}{2})\,+ \zeta(-7, \frac{3}{2})\;
=\,0\ \RA $$
\beq
M\;=\;0.7914,\ \ \ \ \ \ \ \ \ \ V^{(1)}_{fin}(3,4)\;=\;0.0010\,;
\eeq
$$ M\;=\;2.6374,\ \ \ \ \ \ \ \ \ \ V^{(1)}_{fin}(3,4)\;=\;-1.52\
10^{-5}.$$
There are no real solutions for $\,R^2\;\times\;S^2\;$ and
$\,R^3\;\times\;S^3\;$ spaces.  We compare (28) with finite
parts of the effective potential obtained in Ref.7 for
$\,R^4\;\times\;S^N\;$ . The values of $\,V^{(1)}_{fin}\,$
do not agree with those of Ref.7 for even-dimensional spaces
$\,R^4\;\times\;S^2\;$ and $\,R^4\;\times\;S^4\;$. We note that
the finite one-loop potential (29) coincide with the one-loop
potential  obtained in Ref.13 for a conformal scalar field.\\

{\bf 5. The vacuum energy for vector fields} \\

A vector field $\;A_{\mu}\;$ given on $\,R^{m+1}\ \times\  Y\,$
can be expanded as
\beq
A_{\mu}\;=\;(\,A_a\,(x)\,B(y)\;,\;A(x)\,B_\alpha\,(y)\,),
\eeq
where $\,\mu\,=\,0,1,...m+dimY,\ \ a\,=\,0,...m, \ \
\alpha\,=\,1,... dimY,\ \  x,y\,$ are coordinates on
$\,R^{m+1}$ and $\,Y\,$ . For a second order vector
operator $\,\Delta_{\mu\nu}\,=\,-\delta_{\mn}\,\Nk\,+ R_{\mn}$  on $\;S^N\,$
we\
 have
$$ \lambda^{(1)}_n\;=\;n(n\,+\,N\,-1)\,-\,1,\ \ \ \ \ \ \ \ \ \
d^{(1)}_n\;=\;\frac{n(n\,+\,N\,-1)\,(2n\,+\,N\,-1)\,(n\,+\,N\,-3)!}
{ (N\,-2)!\,(n\,+\,1)!}, $$
\beq
\lambda^{(2)}_n\;=\;\lambda_n,\ \ \ \ \ \ \ \ \ \
d^{(2)}_n\;=\;d_n\ \ \ \ \ \ \ \ n\,=\,1,2...
\eeq
where $\;\lambda_n\;$ is given by (13). Starting with
$\;det\,\Delta^{-1/2}_{\mu\nu}\;$ and using (34) we obtain the
one-loop vector potential on $\,R^{m+1}\;\times\;S^N\,$
\beq
V^{(1)}_{v}\;=\;\lim_{s\RA
%% FOLLOWING LINE CANNOT BE BROKEN BEFORE 80 CHAR
1}\;\frac{-\Gamma{((-s-m)/2)}}{2\,(4\pi)^{m/2}}\,((m\,+1)\,\sum_{n}d_{n}{\lambd\
a_{n}}^{(m+s)/2}\;+\;\sum_{n}d^{(1)}_{n}{\lambda^{(1)\;(m+s)/2}_{n})}
\eeq
The calculations of $\;V^{(1)}_{v}\;$ are similar to those for
scalar and spinor fields. Inserting (35) into (36) and using (9)-(11),(18)
after a lengthy but straightforward calculation we get the
following results .

The divergent parts of $\;V^{(1)}_{v}\;$ are
$$ V^{(1)}_{v\;div}(0,3) =\,-0.0625\,\frac{1}{\bar{\epsilon}},
 \ \ \ \ \ \
V^{(1)}_{v\;div}(1,2) =\,-0.0106\,\frac{1}{\bar{\epsilon}},
\ \ \ \ \ \
 V^{(1)}_{v\;div}(1,4) =\,-0.0202\,\frac{1}{\bar{\epsilon}},$$
\beq
 V^{(1)}_{v\;div}(2,3) =\,-0.0033\,\frac{1}{\bar{\epsilon}},
\ \ \ \ \
 V^{(1)}_{v\;div}(3,2) =\,-2.41\ 10^{-4}\,\frac{1}{\bar{\epsilon}},
\ \ \ \ \
 V^{(1)}_{v\;div}(3,4) =\,-0.0012\,\frac{1}{\bar{\epsilon}}.
\eeq
 The finite parts of $\;V^{(1)}_{v}\;$ are
$$ m\ \ \ \ \ \ \ \ \ N\,=\,1\ \ \ \ \ \ \ \ \ \ \ \ N\,=2 \ \ \
\ \ \ \ \ \
\ \ \ \ \ \ \ \ N\,=\,3\ \ \ \ \ \ \ \ \ \ \ \ \ \ \ N = 4$$
$$ 0\ \ \ \ \ \ \ \ \ -0.6667\ \ \ \ \ \ \ \ \ -0.3976\ \ \ \ \
\ \
\ \ \ \ \ \ -0.3198\ \ \ \ \ \ \ \ \ -0.2948 $$
\beq
1\ \ \ \ \ \ \ \ \ -0.0145\ \ \ \ \ \ \ \ \ -0.0467\ \ \ \ \ \ \
\ \ \ \ \ \ -0.0283\ \ \ \ \ \ \ \ \ -0.0455
\eeq
$$ 2\ \ \ \ \ \ \ \ -0.0018\ \ \ \ \ \ \ -0.0023\
\
\ \ \ \ \ \ -0.0047\ \ \ \ \ \ \ -0.0212 $$
 $$ 3\ \ \ \ \ \ \ \ -2.53\ 10^{-4}\ \ \ \ \ \ \ -5.46\ 10^{-4}\
\
\ \ \ \ \ \ 3.92 \ 10^{-4} \ \ \ \ \ \ \ -1.59\ 10^{-4}.$$

We can cancel divergencies and compute $\;V^{(1)}_{v}\;$
choosing the appropriate $\;M\;$. We find for $\;m = 1\;, N = 4\;$
  $$
\frac{1}{4}\,(\frac{9}{4} - M)^3\,+\,\frac{1}{4}\,(\frac{1}{4} -
M)^3\,-\,\frac{3}{16}\,(\frac{9}{4} - M)^2\,-
\frac{27}{16}\,(\frac{1}{4} - M)^2\,+$$
 $$ (3\zeta(-3, \frac{5}{2}) - \frac{27}{4} \zeta(-1,
\frac{5}{2}))(\frac{1}{4} - M)\,+\,
 (2\zeta(-3, \frac{3}{2}) - \frac{1}{2}\,\zeta(-1,\frac{3}{2})\,+
 \zeta(-3, \frac{5}{2}) - $$
 $$ \frac{1}{4}\,\zeta(-1,\frac{5}{2}))(\frac{9}{4} -M) +
\frac{1}{2}\,\zeta(-3, \frac{3}{2}) +
 7\zeta(-3, \frac{5}{2}) -2\zeta(-5, \frac{3}{2}) -4\zeta(-5,
\frac{5}{2})\,= 0$$
\beq
 \RA\ \ \ M\;=\;0.7359,\ \ \ \ \ \ \ \ \ \ V^{(1)}_{v\;fin}(1,4)\;=\;0.0147
\eeq
For $\;m =2\;, N = 3\;$
 $$ M^3\;+\;2\,(M\,-1)^3\;+\;6\,M^2\ =\ 0 $$
\beq
 \RA\ \ \ M\;=\;0.31735,\ \ \ \ \ \ \ \ \ \ V^{(1)}_{v\;fin}(2,3)\;=\;0.0058
\eeq
For $\;m = 3\;, N = 2\;$
 $$ (\frac{1}{4} - M)^3\,+\,(4\zeta(-1, \frac{1}{2})\,+\,2\zeta(-1,
\frac{3}{2}))(\frac{1}{4} - M)^2\,-\;$$
 $$ (8\zeta(-3, \frac{1}{2}) + 4\zeta(-3,
\frac{3}{2}))(\frac{1}{4} - M)\,+\,2\zeta(-5, \frac{3}{2})\,+\,4\zeta(-5,
\frac\
{1}{2})\, =\,
0$$
\beq
 \RA\ \ \ M\;=\;0.1767,\ \ \ \ \ \ \ \ \ \ V^{(1)}_{v\;fin}(3,2)\;=\;6.21\
10^{-4}.
\eeq
There are no real $\;V^{(1)}_{v\;fin}\;$ on
$\,R^2\;\times\;S^2\;$ and $\,R^4\;\times\;S^4\;$.\\

{\bf 6. The one-loop effective potential on $\,R^{m+1}\
\times\;CP^2$ .}\\

Let us start with the harmonic expansion on $CP^2$. For any
homogeneous space $G/H$ a field $\Phi_A$ belonging to
 the irrep $D(H)$ can be expanded as$^7$
 $$
\Phi_A (x)\;=\;\Omega^{-\frac 12}\,\sum_{n, \,i\,q}\;\sqrt
{\frac {d_n}{d_D}} D^{(n)}_{A\;i\;q}\;g_x^{-1}) \phi^{(n)}_{i\,q
}$$
where $V$ is the volume of $G/H$, $d_D={\rm dim}D(H)$.
We summarize over representations $D^{(n)}$ of $G$
which give $D(H)$ after reduction to $H$. Index $i\,$ labels
the multiple components  of $D(H)$ in the branching
$D^{(n)}\downarrow H$, $d_n=$$\dim D^{(n)}$. For scalar fields
on $\;CP^2\;=\;SU(3)/SU(2)\times U(1)\;$ the $D(H)$ is the
trivial representation $\,(1,0)\;$ where the first number in
brackets is the dimension of the $SU(2)$-irrep, and the second one is the
$U(1)$ charge. Using the Yung table method$^{14}$ we find all
representations $D^{(n)}(G)$
giving fixed representation $D(H)$. For the representation
$\,(1,0)\;$ we have
 $$
D(SU(3))=\quad (n,n),\quad n\ge 0
 $$
Here the $\;SU(3)\;$ representations are labelled by the Dynkin
indices . The eigenvalues of the Laplace operator are given by
those of the quadratic Casimir operators of the $\;SU(3)\;$-irreps
$\,(n,n)\;$ . The according degeneracies are the dimensionalities of
the irreps $\,(n,n)\;$
\beq
\lambda^{(1)}_n\;=\;n\,(n\,+\,2),\ \ \ \ \ \ \ \ \ \
d^{(1)}_n\;=\;(n\,+\,1)^3.
\eeq
Substituting (42) in (4) we obtain the following numerical values
of $\;V^{(1)}(m)\;$ :
$$  V^{(1)}(0)\;=\;-0.1783,\ \ \ \ \ \ \ \ \ V^{(1)}(1)\;=\;
-0.0038\,\frac{1}{\bar{\epsilon}}\;-0.0145, $$
\beq  V^{(1)}(2)\;=\;-0.00126,\ \ \ \ \ \ \ \ \ V^{(1)}(3)
\;=\;-1.12\ 10^{-4}\,\frac{1}{\bar{\epsilon}}\,-1.56\ 10^{-4}.
\eeq
We compute the finite one-loop potential
 \beq M\;=\;1.2729,\ \ \ \ \ \ \ \ \
V^{(1)}_{fin}(1)\;=\;9.56\ 10^{-5}.
\eeq
There is no  real parameter $\;M\;$ to cancel the divergence
in (43) when $\;m\,=\,4$.

Now let us consider the vector one-loop potential.
A vector field on $CP^2\;$ has the following $\;SU(2)\times
U(1)\;$ expansion
\beq (2,\,1)\,\oplus \,(2,\,-1)
\eeq
We note that (45) is a real irrep. A set of the $\;SU(3)\;$-irreps
giving (45) under reduction to $\;SU(2)\times U(1)\;$ is
\beq
D\,(SU(3))_{1}\;=\ (n,n),\quad n\ge 1\ \ \ \ \ \ \
D\,(SU(3))_{2}\;=\ (n,\,n+3),\quad n\ge 0
\eeq
For a vector operator $\,\Delta_{\mu\nu}\;$ defined on
$\;CP^2\,$ we have according to (46):
$$ \lambda^{(1)}_n\;=\;n(n\,+\,2)\ \ \ \ \ \ \ \ \ \
d^{(1)}_n\;=\;(n\,+\,1)^3 \ \ \ \ \ \ \ \ n\,=\,1,2...$$
\beq \lambda^{(2)}_n\;=\;n^2\;-\;1/4\ \ \ \ \ \ \ \ \ \
d^{(2)}_n\;=\;n(\,n^2 \,-\;9/4)\ \ \ \ \ \ \ \
n\,=\,5/2\,,7/2...
\eeq
Inserting (47) into (36) and using (9)-(11),(18) we get the following results
:
 $$ V^{(1)}_v(0)\;=\;-0.21968 ,\ \ \ \ \ \ \ \ \
V^{(1)}_v(1)\;=\;-0.00826\,\fr\
ac{1}{\bar{\epsilon}}\, -0.0398,  $$
\beq
  V^{(1)}_v(2)\;=\;-0.0046,\ \ \ \ \ \ \ \ \ V^{(1)}_v(3)\;=\;-4.36\
10^{-4}\frac{1}{\bar{
\epsilon}}\, -0.00165.
\eeq
Removing the divergent part $\,V^{(1)}_{v\,div}(1)\,$ in (48) we find  the
fini\
te one-loop potential
\beq
M\;=\;0.5571,\ \ \ \ \ \ \ \ \ \ V^{(1)}_{v\;fin}\;=\;7.63\ 10^{-4}.
\eeq
There is no the real value $\;M\;$ to obtain the finite
$\;V^{(1)}_{v\;fin}(3)\;$ .\\

{\bf 7. The one-loop potential in electrodynamics on
$\,R^1\;\times\;S^3\;$ , $\,R^2\;\times\;S^2\;$}\\
{\bf and $\,R^3\;\times\;S^1\,$ spaces.}\\

Let us begin with an Abelian gauge theory with fermions in a curved
$\,d\,$-dimensional space-time$^{15}$. The partition function is taken to be
 $$
Z(J,\,\bar{\eta},\,\eta;\,\hat A,\,\hat{ \bar{\Psi}},\,\hat \Psi)\;=
\int\,DA_\mu\,D\bar{c}\,Dc\,D\bar{\Psi}\,D\Psi\,\exp(-\,S\,(\hat A + A) -
S_{f}\,(\hat {\bar{\Psi}}\,+\,\bar{\Psi},\,\hat
\Psi\,+\,\Psi\,)\;$$
\beq
-\,S_{gf}\;-\,S_{gh}\;+\;J^\mu\,A_\mu\;+\;\bar{\Psi}\,\eta\;+\;\bar{\eta}\\
,\Psi\,)
\eeq
where
\beq
%% FOLLOWING LINE CANNOT BE BROKEN BEFORE 80 CHAR
S\,(A)\,+\,S_{f}\,(\bar{\Psi},\,\Psi\,)\,=\,-\,\frac{1}{4}\,\int\,d^dxg^{\frac{\
1}{2}}\,F_{\mn}\,F^{\mn}\,+\,i\,\int\,d^dxg^{\frac{\
1}{2}}\,\bar{\Psi}(D_\mu\,\gamma^\mu\;
 - M_f)\Psi
\eeq
and $J_\mu\;,\;\bar{\eta},\;\eta\;$ are source terms, $\hat
A_{\mu}\;,\,\hat {\bar{\Psi}},\;\hat \Psi\;$ are the arbitrary
background gauge and fermion fields.
 As a gauge fixing term choose
\beq
S_{gf}\;=\;-\,\frac{1}{2}\,\int\,d^dxg^{\frac{\
1}{2}}\,(\hat
 D_\mu\,A^\mu)^2
\eeq
With this choice of gauge-fixing condition, the ghost part of the
action may be seen to be
\beq
S_{gh}\;=\;\int\,d^dxg^{\frac{\
1}{2}}\,\bar{c}\,(-\,\hat D^2\;-\;A^\mu\,\hat
D_\mu\;-\;(\hat D_\mu\,A^\mu)\,c
\eeq
where $\;\bar{c},\,c\;$ are the anticommuting ghost fields$^{16}$,
$\,\hat D_\mu\;$ is the covariant derivative formed using the
background-field gauge connection $\,\hat A_\mu\,$ . In order   to
obtain the one-loop partition function $\ Z^{(1)}\ $ the action
is expanded in powers of the quantum fields $\;A_\mu,\
\bar{\Psi},\;\Psi\;$. The  one-loop effective action
$\;\Gamma^{(1)}(\hat A,\,\hat {\bar{\Psi}},\;\hat \Psi\;)$ which is
connected with $\ Z^{(1)}\ $ by the Legendre transformation
\beq
\Gamma^{(1)}\,\equiv \int\,d^dxg^{\frac{\
1}{2}}\,V^{(1)}\;=\;\log\,Z^{(1)}\;-\;\int\,d^dxg^{\frac{\
1}{2}}\,(J^\mu\,A_\mu\;+\;\bar{\Psi}\,\eta\;+\;\bar{\eta}\,\Psi\,).
\eeq
containes the terms (51)-(53), which  quadratic in the fluctuations.
Substituting (50)-(54) in (55) and after some calculations we obtain
\beq
\Gamma^{(1)} =\frac{1}{2}\log\,det\,\Delta_{\mn} - \log\,det(-\hat
D^2) -\frac{1}{2}\log\,det(-\hat
D^2 -\frac{R}{4}  + M^2_f)
\eeq
where we take $\;\hat {\bar{\Psi}}\,=\,\hat \Psi\,=\,\hat F_{\mn}\,=\,0\;$.
Her\
e
$\,R\,$ is the scalar curvature of  manifold.

The divergent part of $\;\Gamma^{(1)}\;$ has been calculated for
an arbitrary 4-dimensional curved space$^{15}$. It has the following
form (see Eq.(3.48) in Ref.15)
$$
%% FOLLOWING LINE CANNOT BE BROKEN BEFORE 80 CHAR
\Gamma^{(1)}_{div}\;=\;-\;\frac{1}{32\,\pi^2\,\epsilon}\,\int\,d^Nxg^{1/2}\,(-\\
frac{11}{72}\,R^2\;+\;\frac{23}{45}\,R^2_{\mn}\;+
$$
\beq
-\,\frac{19}{360}\,R^2_{\RR}\,-\,2M^4_f\,+\,\frac{1}{3}\,R\,M^2_f).
\eeq

We can easily compute the numerical values of
$\,V^{(1)}_{div}\;$ on 4-dimensional spaces
$\,R^1\,\times\,S^3\,,\,R^3\,\times\,S^1\,$ and $\,R^2\,\times\,
S^2\,$ and compare their with (56).
 Inserting (17),(27),(37) into (55) we find (with $M=0\,$) that
$\,V^{(1)}_{div}\,=\,0\,$ for $\,R^1\,\times\,S^3\,$
and $\,R^3\,\times\,S^1\,$ .
For $\,R^2\,\times\,S^2\,$ we have
\beq
V^{(1)}_{div}\;=\,-\,\frac{1}{40\,\pi\bar{\epsilon}}.
\eeq
 We find the same
values of $\,V^{(1)}_{div}\;$ from (56) (with $\,M_{f}\,=0\,$)
using the values of tensor curvatures on
$\,S^N\,$
 $$
R^2_{\alpha\beta\gamma\delta}\,=\,2N\,(N-1),\ \ \ \ \
R^2_{\alpha\beta}\,=\,N\,(N-1)^2,\ \ \ \ \ \
R^2\,=\,N^2\,(N-1)^2
. $$
According to general prescriptions$^8$ one has to remove the pole
part  (57) to get the finite value of $\,V^{(1)}\,$. However,
the finite potential depends on the scale parameter $\,\mu\,$,
which should be fixed by the application of additional
(theoretical or experimental) conditions.
{}From (17),(28),(38) and (57) we obtain
\beq
V^{(1)}_{fin}\;=\;-0.00796\,\log\mu\rho\,-\,0.02775.
\eeq
To extract physically meaningful finite results, the
effective potential must be renormalized. In our approximation
for massless fields and $\,\hat {F_{\mu\nu}}\,=0\,$
renormalization of coupling constants associated with $\,R^2,\
R^2_{\RR}\ R^2_{\mn}\,$ is enough to render the theory finite.
For two-dimensional space only one of these constants is
independent. Then this coupling constant absorbs the divergent
term in (57). It depends on $\,\mu$ logarithmically so that
$\,V^{(1)}_{ren}\,$ is independent of the normalization scale.
However, renormalization is ill-defined in spaces with
dimensionalities more than four.
On the other hand the divergent term  (57) can be removed by the
choosing the appropriate values of $\,M$.

Now we can compute the finite part of $\,V^{(1)}\,$  . From
(4),(14),(36),(55),\
(12) we have
 $$M_{f}\,=\,0.190794,\ \ \ \ \ \ \ \ \ V^{(1)}_{fin}\;=\;-0.0095 $$
\beq
M_{f}\,=\,-0.5241,\ \ \ \ \ \ \ \ \ V^{(1)}_{fin}\;=\;-0.0481.
\eeq
We also find for $\,R^1\,\times\,S^3\,$
\beq
V^{(1)}_{fin}\;=\;0.1099.
\eeq
 The positive potential (60) gives the repulsive
force in the Einstein static universe ( $\,R^1\,\times\,S^3\,$).
For the $\,R^3\,\times\,S^1\,$ space we have the attractive
Casimir force defined by the potential
$$ V^{(1)}_{fin}\;=\;-0.0017 $$
This result can be easy obtained from Ref.17, where the values
of
$\,V^{(1)}_{sp}\ \ ,\ V^{(1)}_{v}\,$ have been calculated by
different methods.\\

{\bf 8. Conclusion.}\\

In this article we computed the one-loop effective potential for
scalar, spinor and vector fields on
$\,R^{m+1}\,\times\,S^N\,(CP^2)\,$ spaces. We used
zeta-regularization and obtained the finite values of the
effective potential for odd-dimensinal spaces . Since we can expect
the existance of pole terms in $\,V^{(1)}\,$ for even-dimensional
spaces, the finite part $\,V^{(1)}_{fin}\,$ for these spaces
depends on an abitrary scale parameter. This fact may explain
the difference between our results and those of Ref.7 for
$\,R^4\,\times \,S^2,\,S^4\,$ spaces.

As known$^{18}$ the non-zero $\,V^{(1)}_{div}\,$ is
connected with scale symmetry breaking by quantum corrections
and gives rise to a nonvanishing conformal anomaly. The
cancelation of $\,V^{(1)}_{div}\,$ by the choice of $\,M$ may
lead to restoration of scale symmetry . As a result
$\,V^{(1)}_{fin}\,$ in
(21),(22),(25),(29)-(33),(39)-(41),(44),(49),(59)
is independent of $\,\mu\,$ and determined
by topology and geometry of the curved space and fixed value of
$\,M\,$. We note that removing $\,V^{(1)}_{div}\,$ by
introducing $\,M$ leads to a changing of the sign in some values
of $\,V^{(1)}_{fin}\,$  which, in principle , determines the
behaviour of the theory.

The above method can be also applied to non-Abelian gauge fields
with $\,\hat {F_{\mn}}\neq\,0\,$ and tensor fields.\\

{\bf Refereces}\\
1. G.~Plunien, B.Muller and W.Greiner, Phys.Rep.{\bf
134}, 87
(1986)\\
2. H.G.B.Casimir, Proc. K. Ned. Akad. Wet. {\bf 51},
793
(1948)\\
3. J. Schwinger, Phys. Rev. {\bf 174}, 1764 (1968)\\
4. K. A. Milton, Ann. Phys. {\bf 150}, 432 (1983)\\
5. T. Applequist and A.Chodos, Phys. Rev. {\bf D28}, 772
(1983)\\
6. P. Candelas and S. Weinberg, Nucl. Phys. {\bf B237}, 397
(1984)\\
7. D. Birmingham and S. Sen, Ann. Phys. {\bf 172}, 451 (1986)\\
8. S. K. Blau, M. Visser and A. Wipf, Nucl. Phys. {\bf B310},
163 (1988)\\
9. A. Salam and J. Strathdee, Ann. Phys. {\bf 141}, 316 (1982)\\
10. M. Abramowitz and I.A. Stegun, Handbook of mathematical
functions

 (Dover, New York, 1970)\\
11. V.D. Lyakhovsky, N. Shtykov and D.V. Vassilevich,

Lett. Math. Phys. {\bf 21}, 89 (1991)\\
12. M.A. Awada and D.J. Toms, Nucl. Phys. {\bf B245}, 161
(1984)\\
13. L.H. Ford, Phys. Rev. {\bf D 14}, 3304 (1976) \\
14. P.R. Slansky, Phys.Rep.{\bf 79C}, 1 (1981)\\
15. D.J. Toms, Phys. Rev. {\bf D27}, 1803 (1983)\\
16. L.D. Faddeev and V.N. Popov, Phys. Lett. {\bf 25B}, 29
(1967)\\
17. L.H. Ford, Phys. Rev. {\bf D 21}, 933 (1980)\\
18. S.~M. Christensen, Phys. Rev. {\bf D 17}, 946 (1978)

\end{document}